\documentclass[%
 reprint,
 amsmath,amssymb,
 aps,nofootinbib,   
]{revtex4-1}
\usepackage{bm}
\PassOptionsToPackage{linktocpage}{hyperref}
\usepackage[hyperindex,breaklinks]{hyperref}
\usepackage{enumitem}
\usepackage{slashed}

\renewcommand{\theta}{\vartheta}

\renewcommand{\vec}[1]{\ensuremath{\boldsymbol{#1}}}

\newcommand{\bra}[1]{\ensuremath{\left< #1\,\right|}}
\newcommand{\ket}[1]{\ensuremath{\left|\, #1\right>}}

\usepackage{array}
\usepackage{mathtools}

\usepackage{etoolbox}

\begin{document} 

\title{
 Classicalization Clearly:  Quantum  Transition into \\
States 
of Maximal Memory Storage Capacity }

\author{Gia Dvali} 
\affiliation{%
Arnold Sommerfeld Center, Ludwig-Maximilians-Universit\"at, Theresienstra{\ss}e 37, 80333 M\"unchen, Germany, 
}%
 \affiliation{%
Max-Planck-Institut f\"ur Physik, F\"ohringer Ring 6, 80805 M\"unchen, Germany
}%
 \affiliation{%
Center for Cosmology and Particle Physics, Department of Physics, New York University, 726 Broadway, New York, NY 10003, USA
}%

\date{\today}

\begin{abstract}
Classicalization is a phenomenon of redistribution of  energy - initially stored
in few hard quanta - into the high occupation numbers of the soft modes, 
described by a final state 
that is approximately classical.  Using an effective Hamiltonian, we first show 
why the transition amplitudes that increase 
occupation numbers are exponentially suppressed and 
how a very special family of classicalizing theories 
compensates this suppression. 
This is thanks to a large  micro-state entropy generated by the emergent gapless modes
around the final classical state. 
The dressing of the process by the super-soft quanta of these modes compensates 
the exponential suppression of the transition probability. 
Hence, an unsuppressed classicalization takes place 
exclusively into the states of exponentially enhanced memory storage capacity. 
Next, we describe this phenomenon in the language of a quantum neural network, in which the neurons  are represented as interconnected quantum modes with gravity-like negative-energy synaptic connections. 
We show 
that upon an injection of energy in form of a hard quantum stimulus, 
the network reaches the classicalized state of exponentially enhanced memory capacity 
with order one probability. 
We construct a simple model in which the transition results  into classical states that carry an area-law micro-state entropy.
 In this language, a non-Wilsonian UV-completion of the Standard Model via classicalization implies that above cutoff energy the theory operates as a brain network that softens the high energy quanta by bringing itself into the state of a maximal memory capacity.   
It is striking that one and the same underlying mechanism
 can be responsible for seemingly remote phenomena across the disciplines, such as, the formation of classical black holes in elementary particle collisions, the solution  to the Hierarchy Problem 
 via classicalization of the Higgs field, as well as, 
the transition to a maximal memory state in brain networks.   

\end{abstract}


\maketitle

\section{Introduction} 
 It has been hypothesized some time ago \cite{Cl1}  that an alternative to the usual 
 ``Wilsonian" UV-completion of the standard model could exist in form of a self-completion by classicalization.  
 
 The key idea is that - due to a new interaction
 that becomes strong above certain energy scale $\Lambda$ - the theory enters into an 
 {\it effectively-classical} regime in the following sense. In a scattering process above the scale $\Lambda$,  the energy of the initial hard quanta gets redistributed among a large number $N$ of the soft ones \cite{Cl2,2N}.  
 The final quantum state obtained in this way contains a  high occupation number of the soft particles and therefore is {\it approximately classical}.  Hence, the term ``classicalization".  \\

 As explained in \cite{Cl1}, among other motivations, such a mechanism of  UV-completion of the Standard Model would offer a solution to the Hierarchy Problem, as it would 
 classicalize the hard quanta of the Higgs field above the scale $\Lambda$. In this way, the scale $\Lambda$
 would effectively play the role of the cutoff of the 
 quantum theory. For example, a would-be Higgs particle of mass $\gg \Lambda$ must effectively be replaced by a multi-particle classical state.  
 \\

 The question of viability of the Classicalization 
phenomenon has several aspects, all of which cannot 
be addressed in one paper.  We shall focus on one 
particular key aspect: \\

 {\it An unsuppressed quantum transition into a 
 multi-particle classical state. }\\

 The usual physical intuition is telling us  
 that in quantum field theories the  probability of creation of a classical objects of macroscopic occupation number - 
  from an initial state of low occupation number -
 is expected to be exponentially suppressed.  The reason is a high cost in the action for 
 such a transition process.  Thus, semi-classically, the corresponding matrix element is expected to be suppressed by a factor ${\rm e}^{-{S \over \hbar}}$, where $S \gg \hbar $ is an Euclidean action evaluated over a trajectory leading to such a transition.  
Moreover, this expectation is also supported by the experimental 
 data.  For example, we do not observe a production of 
 heavy nuclei in proton-proton collisions at LHC, despite the fact that the center of  mass energy is more than enough for pair-creating such nuclei.  \\

 Thus, in order for {\it classicalization}  to be a viable idea, we will need to understand the two things. 
 First, we need to identify the nature of the exponential suppression
 of the transition amplitudes into the states of high occupation numbers, in generic theories.  Secondly,   
we must understand how the classicalizing theories 
 escape this exponential suppression. 
 What is special about such classical states?  \\ 
 
  An useful qualitative guideline \cite{SELF,Cl2,2N}
 is provided by the well known example of black holes \cite{BH1,BH2,BH3}. 
  Namely,  a simple common sense semi-classical argument indicates that the production of a macroscopic  black hole in the collision of high energy particles should exhibit no exponential suppression.   
   In the microscopic language, the absence of the suppression can be attributed to a high 
entropy of the final state, which provides a required enhancement factor. 
That is, while a production of each black hole micro-state is exponentially suppressed, 
the suppression is compensated by their multiplicity, since they all amount to the same classical macro-state. \\

 This  tells us that an unsuppressed 
 quantum-to-classical transition must represent a transition into a state of {\it exponentially enhanced memory storage capacity and high complexity}. \\
 
  Recently \cite{Gia1,Gia2}, it has been shown  that certain quantum neural networks,
 with {\it gravity-like} synaptic connections,  
exhibit the states of exponentially-enhanced memory storage capacity, in the way somewhat analogous 
  to a black hole situation \cite{Crit,NP}.  The analogy 
 appears even more intriguing, as  some of  
 these neural networks turn out to be isomorphic 
 to a $D$-dimensional quantum field model 
 constructed in \cite{Gia3}. There, the enhanced memory state  translates as a critical state with the emergent {\it gapless}  modes inhabiting an area of a
 $D-1$-dimensional sphere.  Therefore, the corresponding micro-state entropy obeys an 
 area-law, reminiscent  of the 
 Bekenstein entropy of a black hole \cite{Bek}. 
 \\

   Due to the above connections with quantum field theoretic 
   models, such a quantum neural network represents a nice laboratory for testing the idea of classicalization. 
 The gain from such analysis is two-fold. First, the neural networks can teach us about classicalization. Secondly, we learn a possible mechanism by which the brain network can efficiently  bring itself into a state 
 of enhanced memory and pattern recognition capacities in response to an external stimulus. \\
 
  The connection between the quantum neural networks 
  and fields is established by identifying 
  the neural degrees of freedom with the momentum modes  of the field, whereas, the synaptic connections between the neurons are mapped on the 
interactions among the different modes in the Hamiltonian \cite{Gia2}. \\

 In general, visualising the quantum field theory systems, 
 such as the Standard Model, as quantum neural network, 
 allows us to get some fresh perspective on things. 
 For example, we can understand a non-Wilsonian UV-completion \cite{Cl1}  as the ability of the system to move itself 
 into the state of a maximal memory storage capacity.
 In this light, the solution of the Hierarchy Problem 
 by classicalization would mean that the Standard Model 
 works as a remarkable quantum brain network, which above cutoff energy swiftly evolves into a maximal memory state.  
   \\
 
 In addition, the neural network language, can help us to comprehend - at the level of a simple effective 
 Hamiltonian - the essence of the phenomena that are usually blurred by secondary 
 technicalities. For example, since the key focus is on a quantum transition to the classical macro-state of high micro-state entropy,  it can teach us a qualitative lesson about an analogous transition 
 from a two-particle state into a classical  black hole. \\

Below we shall proceed as follows. 
Following \cite{Gia1,Gia2}, we shall first construct a simple Hamiltonian that 
describes a quantum neural network that exhibits a classical state of macroscopic occupation number 
$N$ and an exponentially enhanced memory capacity, due to $N$ emergent gapless modes. 
 We than study the question of transitions to such a 
 classical state 
 from an initial quantum state of sufficiently high energy. 
 In very general terms, we first show why 
 in any sensible quantum theory the transition 
 to each particular micro-state is necessarily exponentially suppressed by a factor ${\rm e}^{-N}$. We then show, how the multiplicity of the micro-states overpowers this suppression. This multiplicity originates 
 from the states of the gapless modes that 
 emerge around the classical state. 
  \\
 
  We establish the following general bound. 
   For an unsuppressed transition into a given classical 
 macro-state, it is necessary that  the number 
${\mathcal N}_{\rm gapless~ species }$  
 of distinct species of the gapless modes emergent in this state, is not less than the occupation number $N$ of the  respective ``constituent" soft modes.       
In other words, the occupation number $N$ of  each soft mode, must be accompanied by at least an equal number 
of the emergent species of the gapless modes: 
  \begin{equation} \label{NSP}
    {\mathcal N}_{\rm gapless~ species } \geqslant N\,.
\end{equation} 
This also implies that the micro-state entropy of the classical state must be higher or equal to the occupation number 
of the soft modes in it: 
  \begin{equation} \label{ENT}
    {\rm Entropy} \geqslant N  
\end{equation} 
 \\
  Finally,  we  briefly review some implications of this phenomenon, for the Standard Model, for Black holes and for quantum brain networks in general.  \\
  
  {\it Note}: Since we work with a well-defined quantum Hamiltonian, 
  one can feel free to ignore the term ``neural network". 
  Our results remain valid for an arbitrary quantum system 
  with a large number of interacting degrees of freedom
 of the type we consider.  
 They show, why the transition to each high occupation number state is exponentially suppressed, and how this
  suppression is compensated by the large micro-state entropy factor within a very special family of the classicalizing theories. \\ 
  
  {\it Note} :  While viewing our models as quantum neural networks \cite{kak} of Hopfield type \cite{Hopfield},
we should point out the key differences in our approach. 
  Our emphasis is not on the learning algorithms, but rather on the energy efficiency of the memory storage.
  The latter we measure by the two criteria: \\
 
 1) Achieving the largest number of patterns that can fit within the maximally-narrow energy gap; 
 
 and 

2) Minimizing the energy cost of the response  
 to an external stimulus. \\
  
  This  pushes us to a very different way of the network usage.  In our cases,  the states of enhanced memory capacity  are the states that support a large number $N$ of the emergent {\it gapless}  neurons.   This boosts the 
memory efficiency:   The number of patterns stored in such states scales as ${\rm e}^N$. \\

The crucial point is that this exponentially large number 
of patterns occupies a microscopically-narrow (or even zero) energy  gap, and moreover, can respond to {\it arbitrarily-soft} stimuli. This is very different from a generic quantum Hopfield network \cite{QNETS}, where the 
same number of patterns would occupy the macroscopically large energy gap, scaling as $N$, and they would be blind to the soft stimuli.  \\
      
   This type of exponential enhancement of the memory storage capacity is precisely what makes the classicalization possible.   
 The system can enjoy the unsuppressed quantum transitions to the states 
  that can store exponentially large number of patterns within maximally narrow energy gaps.

 \section{A model} 
 Following \cite{Gia1,Gia2}, let us first construct a simple prototype model of   
 a quantum neural network, which possesses a state of an exponentially enhanced memory storage capacity. 
 The model consists of the set of neural degrees of freedom represented
 by the  
creation and annihilation operators 
$\hat{a}_k^{\dagger}, \hat{a}_k$, that satisfy the usual    
 oscillator algebra, 
       \begin{equation}  \label{algebra} 
    [\hat{a}_r,\hat{a}_k^{\dagger}] = \delta_{rk}\,, \, \, 
  [\hat{a}_r,\hat{a}_k]  =   [\hat{a}_r^{\dagger},\hat{a}_k^{\dagger}] =0\,,   
    \end{equation} 
 where $k, r = 0,1,2,...,N$ are the labels. 
 We shall be interested in the situation when 
 $N$ is large. 
  The excitation level of a neuron $k$  in a given quantum state 
   is described by an eigenvalue (or an expectation value)  of the corresponding number operator $\hat{n}_k \equiv \hat{a}_k^{\dagger}\hat{a}_k$.
  Thus, the basic states of a neural network are the 
  Fock states described by the ket-vectors 
  $\ket{n_0,n_1,.....,n_N}$ labeled by the excitation levels 
  of the different neurons. \\
   
    A network represented by the  following simple Hamiltonian suffices to capture the essence of the emergence of an enhanced memory state,  
    \begin{equation}
\label{Hamilton} 
 \hat{H} =  \epsilon_0 \hat{n}_0 + 
 \sum_{k\neq 0}  \epsilon_k \hat{n}_k
 -  {\epsilon_0 \over \Lambda}  \hat{n}_0 \sum_{k\neq 0} \epsilon_k \hat{n}_k \,,
 \end{equation} 
 where $\Lambda$ is a parameter of dimensionality 
 of energy.
 The parameters $\epsilon_k,~(k=0,1,...N)$, 
are the threshold energies required for the excitation of individual neurons, in the absence of the interactions.  
 The third term describes the interactions, 
 or equivalently the synaptic connections among the neurons.  The structure of this term is very important.   
Here, we have singled out a neuron $k=0$  
as the {\it master} neuron and have only included the interactions between this neuron and the rest. 
As long as the interactions are sufficiently weak (as we assume), the rest  of the connections play no role in what follows and have not been included explicitly. 
The same applies to the higher order terms that must be included 
in order to make the Hamiltonian bounded from below. 
Nothing in our analysis is affected by the presence of such terms, and therefore, they will be ignored for 
simplicity.  
\\

\begin{figure}
 	\begin{center}
        \includegraphics[width=0.53\textwidth]{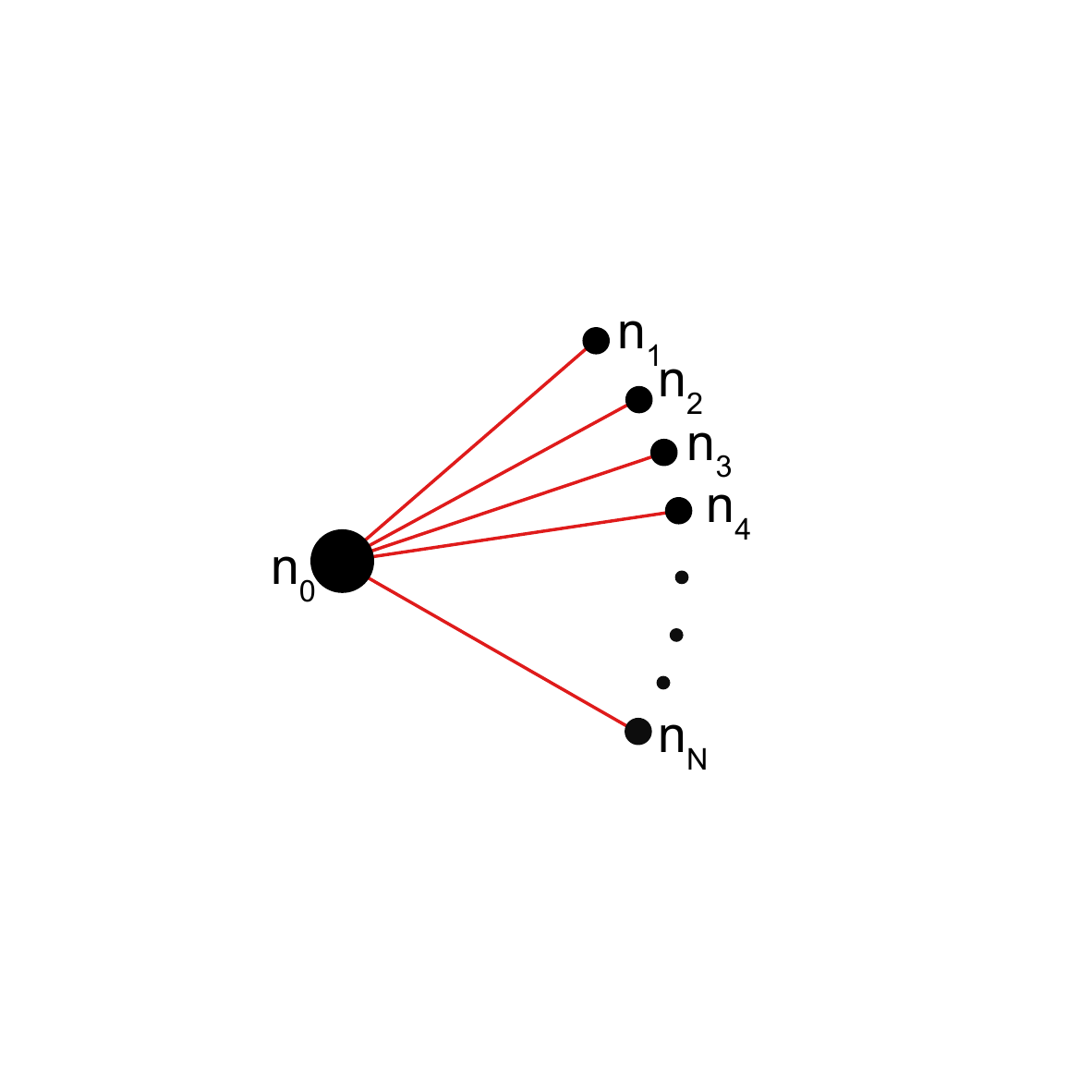}
 		\caption{The network representation of the model (\ref{Hamilton}). The neurons are represented 
		 by the black circles with the size of the circle indicating the excitation level of a given neuron. The connecting red lines represent the excitatory (negative energy) synaptic connections. } 	
\label{QNet}
 	\end{center}
 \end{figure}

  The interaction among $\hat{n}_0$ and  
  $\hat{n}_k$ is universally proportional to the threshold energy  $\epsilon_k$ and the sign is negative.  In this way, the interaction is sort of 
  ``gravity-like".  Since, the interesting critical
  phenomena take place  when the occupation numbers reach the inverse connection strength \cite{Gia1,Gia2},  from now on 
  we shall set
 ${\epsilon_0 \over \Lambda} = {1\over N}$.  
 This makes the coupling strength of the master neuron
 inverse of the number of modes connected  to it.  
  The interaction structure of such a network resonates with 't Hooft's   
large-$N$ limit idea in gauge theories \cite{largeN}.  
 A network representation of the model 
 is given by Fig.(\ref{QNet})
        \\

  The above network represents a simplified version of the models introduced in \cite{Gia1,Gia2}, and reduces the  phenomenon of memory capacity enhancement to its essence.  This essence can be described as follows. \\
 
  The excitation of the master neuron $\hat{n}_0$, 
 due to the negative energy synaptic connections, 
 lowers the excitation thresholds of all the 
 $\hat{n}_{k\neq 0}$ neurons connected to it.  
 That is, on a state $\langle \hat{n}_0 \rangle = n_0$, 
 the  effective excitation threshold of the $k$-neuron is 
 reduced as,
 \begin{equation} \label{Eff}
 \epsilon_k^{\rm (eff)} = \epsilon_k\left (1-{n_0 \over N} \right)\, .  
 \end{equation} 
 Thus, for $n_0 = N$,  all the $k\neq 0$ neurons become effectively {\it gapless}. \\
  
   This means that in the states of these gapless neurons we can store an exponentially large number of patters within an arbitrarily narrow energy gap.  The patterns can be parameterized by the set of $N$ occupation numbers $n_{1},...n_{N}$ and be
stored in the basic states of the form, 
 $\ket{N,n_1,.....,n_N}$. 
 Of course, one can also use the superpositions of the basic states for the pattern storage. \\
 
  All the above basic states are characterized 
   by one and the same macroscopic occupation number 
  of the master neuron, $n_0 = N$.
  If, the numbers
  $n_{k\neq 0}$ are small, then such states can be 
  viewed as the small quantum deviations built around 
 the {\it classical}  ``master" state,  $\ket{N,0,0,.....,0}$.
  In other words, they can be counted as the 
  {\it micro-states} corresponding to the same {\it macro-state} $n_0 = N$.  \\
  
  The legitimacy of the term {\it classical} can be understood in Bogoliubov's sense \cite{bogoliubov}, meaning that  for large $N$ the operators $\hat{a}_0^{\dagger}, \hat{a}_0$ can be effectively replaced by the $c$-numbers.  However, this comment only serves 
  for setting the terminology straight.  Our analysis 
 below will be fully quantum and at no point shall we rely   
 on the Bogoliubov approximation.  \\

   It is important to mention the following.  In the above model, for $n_0=N$, the modes 
    $\hat{n}_{k\neq 0}$ are exactly gapless and therefore the states $\ket{N,n_1,.....,n_N}$ are degenerate for an arbitrary choice of the occupation numbers,  $n_{k\neq 0}$. Nevertheless, only the states 
  with small $n_{k\neq 0}$ are indistinguishable classically, and correspondingly only such states count
 as the representatives of one and the same classical state with $n_0 = N$. 
 For $n_{k\neq 0} \gg 1$, the states correspond to different classical states. Of course, such states 
are equally good for storing the patters and thus play important role in the memory storage. 
However, as we shall see,  in the classicalizing transition in which the system reaches the classical  macro-state of high entropy from some initial low entropy quantum state, only the states with $n_{k\neq 0} \sim  1$
can be reached with an unsuppressed probability. \\

\section{Quantum-to-Classical Transition}

We shall now modify our model in such a way that to allow for a {\it quantum-to-classical} transition.  For this we shall introduce an additional ``hard" degree of freedom
described by the number operator $\hat{n}_b \equiv 
\hat{b}^{\dagger} \hat{b}$ and with a high excitation threshold energy given by 
$\epsilon_b = N\epsilon_0$.  We assume that 
$\hat{b}^{\dagger}, \hat{b}$ operators satisfy the
commutation relations analogous to (\ref{algebra}) and 
commute with all the $\hat{a}$-modes.  \\

 Next, we shall introduce  
a coupling between $b$ and $a$ degrees of freedom
allowing a transition  $\ket{1}_b \rightarrow \ket{N}_a$.  
That is, an input quantum stimulus
 - carried 
by a hard neuron $b$ in its lowest excited state -  is  converted into a classical state of a 
macroscopically excited soft neuron $a$.  As we shall see, 
although the transitions to individual micro-states 
belonging to this classical macro-state are  exponentially suppressed,   
the suppression can be  overpowered, since the classical final state
$\langle \hat{n}_0 \rangle = n_0=N$ exhibits the 
exponentially enhanced memory storage capacity.\\

Thus, the modes emerging around the ``master" state
$n_0=N$, can be classified in three categories,  according to the level of their softness. These are: \\

1) The {\it hard} $b$-mode, with the gap $N\epsilon_0$; 

2) The {\it soft} $a_0$-mode, with the gap $\epsilon_0$; and 

3) The 
{\it super-soft}  $a_{k\neq 0}$-modes, which are essentially gapless. \\ 

The latter 
modes play an absolutely crucial role in making
the classicalizing transition probable: 
It is the summation over the final states that are ``dressed" by the different occupation numbers of these
gapless modes that compensates for the exponential suppression factor.  Clearly, in the same time, this goes hand in hand with the enhancement of  the memory capacity, since an exponentially large number of  patterns are stored in the states of the gapless modes.   
\\     

The above-described classicalizing transitions are generated by the following Hamiltonian,
    \begin{equation}
\label{H2} 
 \hat{H} =  \hat{H}_{\rm numb} \, + \, \hat{H}_{\rm tr} \,,  
  \end{equation} 
where $\hat{H}_{\rm numb}$ denotes the part that conserves the individual occupation numbers of each mode, 
   \begin{equation}
\label{Hnumb} 
 \hat{H}_{\rm numb} =  N\epsilon_0 \hat{n}_b \, + \, \epsilon_0 \hat{n}_0 + 
 \sum_{k\neq 0}  \epsilon_k \left ( 1
 -  {\hat{n}_0 \over N}  \right ) \hat{n}_k \,, 
 \end{equation} 
 whereas $\hat{H}_{\rm tr}$ generates the number-non-conserving transitions. It has the following form 
     \begin{equation}
\label{Htr} 
 \hat{H}_{tr}  \, = \,  \hat{b}^{\dagger}  \hat{a}_0^N \sum_{n_1,n_2,...n_N}  g_{n_1...n_N} \hat{a}_1^{n_1}\hat{a}_2^{n_2}...\hat{a}_N^{n_N} \,  + \, {\rm h.c.}\, ,
 \end{equation} 
where the summation is taken over all possible sets 
of integers $n_1,...n_N$.  \\

Notice, the above interaction must be viewed as  
an effective Hamiltonian describing the transition. 
At the level of  a more fundamental quantum field 
theory it can emerge as a fundamental vertex or be a result of 
summation of an infinite set of Feynman diagrams, 
as, e.g., in \cite{2N}. The power of the analysis at the level of an effective  Hamiltonian is that it uncovers the generic constraints that such interaction must satisfy, regardless of its underlying origin.  \\

We are interested in the time-evolution generated by the above Hamiltonian 
starting from an initial state $\ket{in} \equiv  \ket{1}_b
\times \ket{0,0,...0}_a$, in which the hard nauron
$b$ is excited to the first level, whereas the $a$-neurons are unexcited.  
 An each member of the sum in (\ref{Htr}), after acting on the initial state $\ket{in}$, converts it into the following state, 
 \begin{equation} \label{transit}
  \sqrt{N! n_1!..n_N!}g_{n_1...n_N} 
 \ket{N}_{n_1...n_N} \,,
  \end{equation}  
 where we have introduced a notation, 
 $\ket{N}_{n_1...n_N} \, \equiv \, 
 \ket{0}_b\times \ket{N,n_1,..n_N}_a$.   
This conversion results into the following transition matrix elements, 
 \begin{equation} \label{matrixA}
 |\bra{in} \hat{H}_{\rm tr} \ket{N}_{n_1...n_N}|^2 =  N! n_1!..n_N! |g_{n_1...n_N}|^2 \, .
 \end{equation} 
 Correspondingly, the probability of the transition to one of  the classical states with $n_0=N$, can be obtained by summing over the numbers $n_1, ...,n_N$. \\
 
 Here we arrive to a very important point.  
 Naively, one may think that the above sum can be made 
 arbitrarily large at the expense of adjusting the parameters  $g_{n_1...n_N}$, as well as, by
 making use of the multiplicity of states.
 However, things are not so cheap, because the quantities $g_{n_1...n_N}$
 are subject to the severe consistency constraints. \\
 
   The point is that in order to have a well-defined transition process,  both the initial and the final states 
   must be the approximate eigenstates of the Hamiltonian.
         
   This implies that the  off-diagonal (number-non-conserving) terms in the 
   Hamiltonian must be 
   sub-leading to the number conserving part $\hat{H}_{\rm numb}$ over the {\it entire}  
   part of the Hilbert space that can be probed during this
    evolution of the system.  In particular, this must be true for all the states for which the expectation values 
  of the occupation numbers are centered around the 
 initial and final values involved in the transition. 
 Therefore,  this requirement must be satisfied on all the coherent states $\ket{C}$ with the relevant values 
of the  mean occupation numbers, since such states  represent the  Poissonian distributions centered around the mean occupation numbers $n$ with the variances given by 
 $\Delta n^2 \sim n$. \\

  Thus,  consider a coherent state $\ket{C}$, with the following mean occupation numbers: 
  \begin{equation} \label{COH}
 \bra{C} \hat{n}_b\ket{C} = 1,~ 
 \bra{C} \hat{n}_0 \ket{C} = N,~
 \bra{C}\hat{n}_k \ket{C} = n_k \,.
 \end{equation}   
 We require that on such a state the expectation values of all off-diagonal term must be less than the expectation values of the diagonal ones. 
 Taking the expectation values and remembering that 
 the coherent state is an eigenstate of all the annihilation operators with the eigenvalues given by the square-roots of the corresponding mean-occupation numbers, we obtain the following constraint  
  \begin{equation} \label{cond}
 |g_{n_1...n_N}|  \lesssim \lambda (\epsilon_0N) \, N^{-{N\over 2}} n_1^{-{n_2\over 2}} ...
n_N^{-{n_N\over 2}} \,. 
\end{equation}
Here the factor $\epsilon_0N$  comes from the expectation value of $\hat{H}_{\rm numb}$ and  
$\lambda$ is a small dimensionless parameter that 
controls the relevant strength of off-diagonal expectation 
values.  The precise value is unimportant for our conclusions.
For example, taking, $\lambda = {1 \over N\sqrt{N}}$
implies that the relative strength is less than ${\epsilon_0 \over \sqrt{N}}$. Below, for definiteness, 
we shall make this choice, although it is not unique
and has no important meaning  at this level. \\

The equation (\ref{cond})  represents an extremely powerful constraint which reveals
a lot of information about the nature of the suppression of quantum transition to the states with high occupation numbers. \\

In order to see this, let us examine the cases of 
small and large occupation numbers $n_{k\neq0}$ separately.   Consider first  the transition into a particular micro-state with $n_{k\neq0} \lesssim1$. 
From (\ref{cond}) it is clear that the corresponding 
coefficients are suppressed as $|g_{n_1...n_N}|  \lesssim {\epsilon_0 \over \sqrt{N}}  N^{-{N\over 2}}$.  Then, evaluating the matrix element  (\ref{matrixA}) we obtain,
\begin{equation} \label{matrixB}
 |\bra{in} \hat{H}_{\rm tr} \ket{N}_{n_1...n_N}|^2 \sim  
  {\epsilon_0^2 \over N} N! N^{-N}\, , 
 \end{equation} 
where we took into account that
$n_{k\neq0} \lesssim1$. Using Stirling formula for 
large $N$ we get, 
\begin{equation} \label{matrix1}
 |\bra{in} \hat{H}_{\rm tr} \ket{N}_{n_1...n_N}|^2 \sim  
 \epsilon_0^2 {1 \over \sqrt{N}} {\rm e}^{-N}\, .  
 \end{equation} 
Thus, from very general considerations, we have obtained  an anticipated exponential suppression factor of the matrix element describing a quantum 
transition into a classical state of some large 
occupation $N$. 
This also represents a nice consistency check of our analysis. \\ 

  The above suppression factor is due to a large occupation number of $n_0$-mode in the final state. 
 However, in the present model  this state is accompanied by the emergence of the  $N$ gapless 
$\hat{n}_k$-modes,  which results into an exponential enhancement of the number of micro-states 
with $n_0=N$. \\
 
  As a result,  the above suppression factor is matched by the multiplicity of states with $n_{k\neq0} \lesssim1$, 
which is approximately ${\rm e}^{N}$. Summing over all such micro-states, we obtain, 
in the leading order in $g$-expansion, the following transition probability to a classical macro-state, per time $t$,  
\begin{eqnarray} 
\label{matrix}
 && P_{b\rightarrow Na_{0}} = \\
&& = \sum_{n_{k\neq0} \lesssim1}
  |\bra{in} {t \over \hbar} \hat{H}_{\rm tr} \ket{N}_{n_1...n_N}|^2 \sim  
   {t^2\epsilon_0^2 \over \hbar^2} {1 \over \sqrt{N}}\, . \nonumber   
 \end{eqnarray}

 Thus, summarizing:  We observe that the transition probability to the master macro-state of enhanced memory capacity is exponentially more probable than
a would-be transition into an analogous state in absence 
of the memory enhancement. \\

 This enhancement is  due to a large micro-state entropy provided by the 
modes $\hat{n}_{k\neq 1}$ that are rendered gapless by the high occupation number of the master neuron
$\hat{n}_0$.  Indeed, as it is clear from (\ref{matrix1}),  
in the absence of the gapless modes, the transition to a state $n_0 = N$, would be exponentially unlikely. 
It is the presence of the gapless modes 
that provides the necessary compensating factor 
${\rm e}^{N}$, which makes this transition highly probably. \\

 One may wonder whether we would have gotten an  even more enhanced transition rate, had we taken into account also the transitions into the states 
 with large values of  $n_{k\neq0}$-s. 
 After all, the multiplicity of states with $n_{k\neq0} \gg 1$ is much higher than the one of the states with  $n_{k\neq0} \sim 1$.
 For example, the number of all possible states
with  $n_{k\neq0} < d$, where $d$ is some 
integer, scales as  $\sim d^{N}$.  This may create a false impression  that by increasing $d$ we can make the transition more and more probable.  \\

In reality, this is not the case and the reason is the constraint (\ref{cond}).  Evaluating the transition matrix element subject to this constraint, we get:
\begin{equation} \label{matrix2}
 |\bra{in} \hat{H}_{\rm tr} \ket{N}_{n_1...n_N}|^2 \sim  
   N! n_1!...n_N!  N^{-N}n_1^{-n_1}...n_N^{-n_N}\,,   
 \end{equation} 
 which after using Stirling formula gives: 
\begin{equation} \label{matrix2}
 |\bra{in} \hat{H}_{\rm tr} \ket{N}_{n_1...n_N}|^2 \sim  
  \sqrt{Nn_1..n_N}\, 
 {\rm e}^{-(N + n_1 +...+n_N)}\, .  
 \end{equation} 
In both expressions we have omitted an 
unimportant factor $\lambda \epsilon_0 N$.
From the expression (\ref{matrix2}) it is obvious that a high  multiplicity of states with large $n_{k\neq 1}$-s is unable to compensate the respective  
exponential suppression. Indeed, the number 
of such states can at best provide an enhancement 
factor $\sim {\rm e}^{Nln(d)}$, which for $d \gg 1$ is absolutely 
powerless against the suppression factor 
in (\ref{matrix2}),
which scales as 
$\sim {\rm e}^{- N(1 + d)}$. 
We thus conclude that the compensation of the  
exponential suppression factor takes place 
exceptionally for the transitions into the states with small $d$. \\  

  This result makes perfect physical sense. Indeed, 
  the states with large occupation numbers 
  $n_{k\neq 0}$,  are {\it classically-distinguishable} 
  from the master state $\ket{N}_{0...0}$. 
  So, despite of not being separated by a large energy gap, 
 such states nevertheless correspond to  {\it different 
 macro-states}. However, unlike the master state 
 $\ket{N}_{0...0}$, they lack their own ``entourages" of 
 the exponentially large number of the quantum micro-states.  The reason is that the increase of the occupation 
 numbers $n_{k\neq 0}$ is not accompanied by
 an emergence of any new gapless modes.  But, such modes are necessary for providing the further exponential increases of the numbers of micro-states by the respective  factors 
 ${\rm e}^{n_k}$.  Correspondingly, the exponential suppressions of transitions to large-$n_{k\neq0}$ states remain uncompensated. \\ 
 
  Thus, we observe that in a classicalizing transition the system mainly explores the micro-states that are not distinguishable classically from the master state $\ket{N}_{0...0}$, i.e., the micro-states that are obtained 
 from the latter state via small quantum excursions. 
 The more distant classical states remain unexplored. \\
 
 \section{Varying number of soft quanta} 
 
  We are now prepared for the next step, in which 
  we can allow also the transitions to the classical 
  states in which the number of soft quanta $n_0$ can be  different from the critical value $N$. 
 For this, we modify the transition Hamiltonian 
 (\ref{Htr}) as,  
      \begin{equation}
\label{Htr1} 
 \hat{H}_{tr}  \, = \,  \hat{b}^{\dagger}  \sum_{n_0,n_1,...n_N}  g_{n_0n_1...n_N} 
 \hat{a}_0^{n_0} \hat{a}_1^{n_1}...\hat{a}_N^{n_N} \,  + \, {\rm h.c.}\, ,
 \end{equation} 
 where the parameters obey the respective conditions 
of the type (\ref{cond}).  
This leads to processes in which creation 
of $n_0$ soft quanta is accompanied by certain distributions of the rest, $n_{k\neq 0}$.   
As it is clear from (\ref{Eff}), for the special case of 
$n_0\simeq N$, the rest of the modes  become super-soft 
 (i.e., nearly-gapless) and the process can be presented  as: 
 \begin{equation}
 1_{\rm hard}  \rightarrow n_0 + n_{\rm supersoft} \, ,
 \end{equation}
 where  $n_{\rm supersoft}= \sum_{k \neq 0} n_k$.\\
  
 Due to the previous discussion,  the probability of classicalization for large  $n_{0}$ effectively factorizes,  
   \begin{equation}
  P_{class} \,  = \,  
   P_{1\rightarrow n_0} \,  P_{super-soft} (n_0, \epsilon) \, ,
 \end{equation}
where $P_{1\rightarrow n_0} \sim {\rm e}^{-n_0}$ represent the transition 
probability into $n_0$ soft quanta, without accompanying 
super-soft quanta, whereas,  
$P_{super-soft} (n_0, \epsilon)$ takes into the account the ``dressing" of the final state 
by populating it  with the super-soft quanta of the gapless modes. 
It includes summation over all the micro-states 
$\ket{n_0}_{n_1,...,n_N}$
that fit within some fixed energy gap $\epsilon$. 
That is, these are the micro-states 
that satisfy the constraint $\sum_{k\neq 0}  n_k\epsilon_k\left (1- 
{n_0 \over N} \right ) < \epsilon$. 
Due to this, the quantity $P_{super-soft} (n_0, \epsilon)$ is a function of both $n_0$ and of
$\epsilon$. \\

The choice of $\epsilon$ is dictated by 
how much energy uncertainty one is willing to attribute 
in defining the classical macro-state $\ket{n_0}_{0,..0}$.
 An obvious requirement is that this uncertainty must vanish in the exact classical limit,  $n_0 \rightarrow 0$.   
 In our case, the physically justified choice of 
$\epsilon$ is to take it to be less or equal to an elementary gap
$\epsilon_0$.  Then,  given $\epsilon_k \geqslant  \epsilon_0$,
we have the following situation. \\

 First,  for  $n_0 = N$ all the $k\neq 0$ modes are gapless. 
 Since, as explained above,  the states with  $n_k \gg 1$ contribute the exponentially suppressed 
 weights into the probability, the sum is effectively cut off 
 above $n_{k\neq 0} \sim 1$, so that 
 $P_{super-soft} (n_0 =N, \epsilon_0) \sim {\rm e^N}$. \\ 
  
  On the other hand, for $n_0 \neq N$, the number 
  of the states that can fit within the gap $\epsilon$ rapidly diminishes as a function of $|n_0 - N|$. 
   Correspondingly, the function is exponentially sharply peaked at $n_0 = N$ and rapidly falls-off away from it.  Thus, for  $\epsilon \lesssim \epsilon_0$, we can 
   approximate, 
   \begin{equation} 
   P_{super-soft} (n_0, \epsilon) = 
   \delta_{n_0N} {\rm e^N}\,, 
  \end{equation}   
  where $\delta_{n_0N} = \begin{cases}
 0   & \text{for},\,  n_0  \neq N  \\  
  1     & \text{for}, \, n_0 = N
\end{cases} \, \, $  is Kronecker delta.  
   Correspondingly, the
 transition probability is peaked around
 $n_0 = N$, where it is given by
 (\ref{matrix}).  
  
 \section{Classicalization for Arbitrary $N$ and 
 Energy-Dependence of Entropy} 
 
 By now, the following is clear from the previous discussion. The necessary condition for an unsuppressed classicalization  into 
a macro-state $\ket{N}_{a_{0_N}}$, in which the occupation number of a soft mode $a_{0_N}$  is equal to 
$N$,  is that the micro-state entropy of $\ket{N}_{a_{0_N}}$ is larger or equal to $N$, in accordance with the  bound (\ref{ENT}). 
   That is, the number 
  of species of the emergent gapless modes in this state
 is bounded from below by $N$, as indicated in (\ref{NSP}).  \\
 
 This statement fully generalizes to the case when 
 the classical macro-state represents a distribution 
 over the occupation numbers of several different soft modes: 
 \begin{equation} \label{product}
 \ket{class} = \prod_{N} \otimes \ket{N}_{a_{0_N}}\,,
 \end{equation}
 where $\otimes$ stands for a tensor product. 
 In such a case, a macroscopic occupation number 
 $N$ of each soft mode $a_{0_N}$ must be accompanied by $N$ emergent gapless (i.e., super-soft) modes $a_{k_N}$, where $k_N = 1,2,...N$. 
 An each set of the gapless modes will contribute a corresponding factor $N$ in the 
micro-state entropy. Thus, the total micro-state entropy of the state $\ket{class}$ is equal to a total number of the emergent gapless modes:
\begin{equation} \label{summation} 
 {\rm Entropy } = \sum_{a_{0_N}} N.
 \end{equation} 
 In the same time, the relation between the  
 micro-state entropy and the energy of the state 
  is not uniquely fixed 
 by the requirement of classicalization. \\ 
 
 In order to make the above statements clear, 
 below we give an example of the network 
 that allows for classicalizing transitions to the 
 macro-states with arbitrarily-high occupation number $N$ and varying energy $E_N$.  For this, we shall endow both the 
soft master mode as well as the hard modes, with an index $N$, as we did above.  A given master neuron $a_{0_N}$, when 
excited to a corresponding critical level $N$, renders a set of 
$N$ neuron species, $a_{k_N},~k_N = 1,2, ..N$, gapless.  In this way, the system delivers an infinite sequence of macro-states, $\ket{N}_{a_{0_N}}$,
each accompanied by $N$ emergent gapless 
modes $a_{k_N}$  and a respective micro-state entropy equal to $N$.   All such states saturate the bounds 
(\ref{NSP}) and (\ref{ENT}). \\ 

 A simple Hamiltonian that accomplishes this goal has the following form,
 \begin{eqnarray} \label{HHH}  
  && \hat{H}  =  \sum_N N\epsilon_{0_N} \hat{b}^{\dagger}_N \hat{b}_N  +   \sum_N \epsilon_{0_N}  \hat{a}^{\dagger}_{0_N} \hat{a}_{0_N} + \\
  &&   \sum_N \sum_{k_N =1}^{N} \epsilon_{k_N} \left ( 1 - {\hat{a}^{\dagger}_{0_N} \hat{a}_{0_N} \over N}\right )  \hat{a}^{\dagger}_{k_N} \hat{a}_{k_N} + \nonumber\\
  &&   \sum_N\hat{b}_N^{\dagger}  \hat{a}_{0_N}^N \sum_{n_1,n_2,...n_N}  g_{n_1...n_N}^{(N)} \hat{a}_{1_N}^{n_1}\hat{a}_{2_N}^{n_2}...\hat{a}_{N_N}^{n_N} \,  + \, {\rm h.c.}\, .  \nonumber 
   \end{eqnarray} 
 Notice, the different species of the soft master neurons 
 $a_{0_N}$ are now allowed to have different energy gaps $\epsilon_{0_N}$. Correspondingly, we have introduced a spectrum of the hard neuron species, 
 $b_{N}$, with their threshold energies given by 
 $E_N=N\epsilon_{0_N}$. \\
 
   In this system, an input excitation  $\ket{1}_{b_N}$, of energy $E_N = N\epsilon_{0_N}$ and zero entropy, classicalizes into a macro-state  $\ket{N}_{a_{0_N}}$ of the same energy and the entropy given by $N$.  
 Obviously, the transition probability is given by (\ref{matrix}), where $\epsilon_0$ must be
 replaced by  $\epsilon_{0_N}$. \\  
     
It is clear that  the relation  between energy 
and entropy is not uniquely fixed by the requirement of classicalization.
It appears that for various relations, the system can classicalize equally well. \\

  For example, consider the following two choices of
 the soft energy gap $\epsilon_{0_N}$ as function of $N$. 
    
  In the first case we take,   $\epsilon_{0_N} = \epsilon_0 = {\rm constant}$, whereas in the second case, we take  
   $\epsilon_{0_N} = {\Lambda \over \sqrt{N}}$, where 
  $\Lambda$ is some fundamental scale. 
 For both choices, the micro-state entropy of the macro-state $\ket{N}_{a_{0_N}}$ is equal to $N$. 
 However, the energy dependences in the two cases are very different. 
 In the first case, the entropy scales proportional to energy, 
  ${\rm Entr} =N = {E_N \over \epsilon_0}$.  
 In contrast,  in the second case it scales as
 the square of the energy $N = {E_N^2 \over \Lambda^2}$. Notice, the latter scaling is very similar 
  to the relation between the  black hole entropy and  its 
  energy  in $D=3$ space dimensions. \\
  
 Despite the above difference,  in both cases the classicalizing  transition from the state $\ket{1}_{b_N}$
 to the state $\ket{N}_{a_{0_N}}$ 
 takes place without any exponential suppression, 
since  in both cases the number of the emergent gapless species is equal to $N$. This is all 
that is needed for the compensation of the suppression factor ${\rm e}^{-N}$ of the transition amplitude.

 \section{Implications} 
 
  By implementing the construction of \cite{Gia1,Gia2}, we have obtained a simplest quantum neural network which exhibits a state of exponentially enhanced memory capacity due to emergent gapless neurons.  
  We then showed that when a stimulus is injected 
  in the network in form of an elementary
 excitation of some 
  high-energy neuron $\hat{b}$, the system reaches 
  the state of enhanced memory capacity 
  with order one probability. 
 In this process the energy of the hard stimulus 
gets converted into a macroscopically high excitation
of a soft master neuron  $\hat{a}_0$. Schematically, this transition can be described as, 
   \begin{equation} \label{CLT}
\ket{{\rm Quant},n_b =1}_{{\rm Ent}=0}
\rightarrow \ket{{\rm Class}, n_{a_0} =N}_{{\rm Ent}=N} \,,
\end{equation} 
where the subscripts indicates the value of the  micro-state entropy of the initial and final states. 
\\

   Reduced to its bare essentials  the phenomenon
  can be described as follows. A system with many degrees of freedom with negative energy connections
  can exhibit the enhanced memory state in which 
  the high occupation number of one of the soft modes 
 renders the set of other modes gapless \cite{Gia1,Gia2,Gia3}. 
 This enhancement makes the transition to such a
 high occupation number state maximally probable, even when initially the entire energy of the system is concentrated in a single quantum. \\
 
   Such a system defies the ordinary intuition about the 
 exponential suppression of the quantum-to-classical 
 transitions, due to the fact that the resulting classical state is accompanied by  an exponential enhancement of memory storage capacity and corresponding large micro-state entropy. \\ 
 
 Below we shall briefly discuss some implications of our 
 results. 
 
 \subsection{Quantum Brain Networks} 
 
 One obvious implication of our results is for  understanding an accessibility of nearly-classical  states of enhanced memory storage capacity in 
 quantum brain networks, under the influence of the external quantum stimuli.   
As we explained, in such networks the 
existence of energy-efficient memory capacity  
is linked with the emergence of the gapless neurons
\cite{Gia1,Gia2}. 
The latter property also ensures the sensitivity to an arbitrarily soft external stimuli.  
 Correspondingly,  the property of classicalization is intrinsically linked with the ability of the network to reach 
 the highest possible memory state, in which it will also become sensitive to maximally soft stimuli.  \\
 
  To put it shortly:  {\it A hard stimulus prompts the brain network to become responsive to the exponentially large variety of the super-soft input patterns. }

\subsection{Classicalization in Black Hole Formation in High Energy Particle Collisions} 
 
   It is a well-accepted idea \cite{BH1,BH2,BH3} that black holes are expected to form with order one probability in the collision of quantum particles of sufficiently high center of mass energy.  
    
   The qualitative argument is rather simple: Such a system should lead to a black hole formation once the initial center of mass energy becomes concentrated 
 within the corresponding gravitational radius. 
 This phenomenon represents an example  
 of classicalization, since it describes a transition from an initial quantum state of small occupation number 
 of hard quanta into a classical final state.  
   However, the precise microscopic mechanism behind this process in still unknown. 
    The purpose of the present note is not to offer one, 
    but rather to point out a remarkable analogy. \\
    
 Namely, in its essence,  
 the classicalization process of a black hole formation
 is  strikingly similar to the classicalization process in the neural network described above.    
    Indeed, the black hole formation, say, in two-particle collision,    
    represents a transition from an initial quantum state
- in which the entire energy of the system is carried by
 two hard degrees of freedom - into a classical state (i.e., a macroscopic black hole) of enormous memory storage capacity. 
This is exactly what is happening in the neural network presented above. 
In both cases an initial quantum state of 
low memory storage capacity evolves into 
a classical state  of sharply enhanced memory storage capacity:  
\begin{equation} \label{CLT}
\ket{{\rm Quant}}_{{\rm Low~Entropy}}
\rightarrow \ket{{\rm Class}}_{{\rm High ~Entropy}}
\end{equation} 
Of course, it would be too naive to suggest that our model captures all the aspects of gravitational physics. However,  
it may very well be capturing some key aspects of the 
classicalization in the black hole formation, by highlighting the role of the memory capacity enhancement in this process. \\
  
However, there is more to it. 
 In fact, the computations of \cite{2N,2ND},
may serve as strong supporting evidence for such a connection. In these papers, the  relevant quantum gravity process for the black hole formation was identified  as $2 \rightarrow N$ transition, in which  a two initial gravitons of high center of mass energy  
scatter into $N$ equally soft ones.  Especially interesting things happen in the regime when the number $N$ of the soft gravitons is equal to the inverse strength of their quantum gravitational coupling. 
As it is easy to see, the same number  coincides with the entropy of a black hole of the size given by the de Broglie wavelength 
of the soft gravitons.  Equivalently, $N$ 
is the entropy of a black hole with the mass
equal to the center of mass energy of the initial two gravitons. 
This striking coincidence with the qualitative properties of classicalization in the gravity-like neural network considered above, should already ring a bell.  
\\  
 
 In fact, 
 according to the hypothesis of \cite{NP}, such a soft $N$-graviton state
 is at the quantum critical point \cite{Crit} and 
 represents an approximate  (in large $N$) description of 
 a black hole {\it macro-state}. Due to its quantum criticality, 
 there emerges a number $N$ of nearly-gapless super-soft modes. 
 These were assumed to be responsible for the origin of black hole entropy.       
  In this description, the black hole micro-states represent the states of $N$-soft gravitons, accompanied by the distributions of the super-soft ones. The latter gravitons 
  represents the excitations of the gapless modes. \\  

Now, both, the string-theoretic as well as the 
quantum field-theoretic computations of \cite{2N} show that such a scattering  process indeed delivers a suppression factor $ \sim {\rm e}^{-N}$ 
 for the probability of producing a particular
 micro-state. 
Notice, this is strikingly similar to the analogous factor 
in (\ref{matrix1}). 
Moreover, this is precisely the factor that is compensated by the summation over the black hole micro-states, produced by the super-soft modes.  
 \\
 
  In addition, the computation of 
 Addazi, Bianchi and Veneziano \cite{2ND} 
  suggests that the summation over micro-states  
 may have a diagrammatic counterpart in form of the ``dressing" of the $2\rightarrow N$ process by both virtual and real super-soft gravitons.  
\\

 So, if the above picture is even remotely correct,  
 it follows that a most naive gravity-like quantum network captures the qualitative properties of
 a classicalization process in black hole creation in 
 particle collisions. \\

In order to make the analogy with a black hole formation
even sharper,  let us note that using the recipe of 
\cite{Gia2,Gia3}, it is rather straightforward to modify
 the model (\ref{HHH}) in such a way that the
micro-state entropy of the enhanced memory 
classical state shall 
obey an area-law, reminiscent of black hole's  Bekenstein 
entropy \cite{Bek}.  
 For this, we need to impose a  spherical 
symmetry on the neural degrees of freedom, by 
visualizing them as the angular momentum modes 
of some ``parent" field  living in $D$-dimensions \cite{Gia2}.  In particular, the index $k_N$ in 
(\ref{HHH}), must be understood as the label 
of spherical harmonics of a level of degeneracy $N$. \\   

This mapping allows to attribute a well-defined meaning of a local geometry to such an intrinsically
non-local  system as the neural network.  
 The example of a network exhibiting 
an area-law entropy (for the case of angular invariance in $D=3$) is given by the following Hamiltonian: 
 \begin{eqnarray} \label{AH}  
  && \hat{H}  =  \sum_N N \epsilon_{0_N}\hat{b}^{\dagger}_N \hat{b}_N  +   \sum_N \epsilon_{0_N} \hat{a}^{\dagger}_{0_N} \hat{a}_{0_N} + \\
  &&   \sum_N \epsilon_{N} \left ( 1 - {\hat{a}^{\dagger}_{0_N} \hat{a}_{0_N} \over N}\right ) \sum_{m,l} \hat{a}^{\dagger}_{slm} \hat{a}_{slm} + \nonumber\\
  &&   \sum_N\hat{b}_N^{\dagger}  \hat{a}_{0_N}^N \sum_{n_1,n_2,...n_N}  g_{n_1...n_N}^{(N)} \hat{a}_{sl_1m_1}^{n_1}\hat{a}_{sl_2m_2}^{n_2}...\hat{a}_{sl_Nm_N}^{n_N} \,  + \, {\rm h.c.}\, .  \nonumber 
   \end{eqnarray} 
This Hamiltonian has a structure very similar 
to (\ref{HHH}) except it incorporates the spherical symmetry.   Here,  $s=0,1,...\infty$ is an integer that labels    
 the level of angular harmonics 
 on a three-sphere.  
For a given $s$, the integers $m,l$ label 
the usual angular harmonics on $S_2$ and take the values  $|m| \leqslant l \leqslant s $. 
Thus, for fixed $s$, the sum over 
$l,m$ is the sum over the spherical harmonics
of a $2$-sphere belonging to the level 
$s$ of an angular harmonics of a $3$-sphere. 
The summation is subject to an obvious constraint of the angular momentum conservation.  Since, the degeneracy of the level $s$
is $(s+1)^2$, we constrain 
$N$ to take the values $N = (s+1)^2$.  
The couplings  $g_{n_1...n_N}^{(N)}$, 
as before, obey the constraints (\ref{cond}). \\ 

Notice, we do not need to require the full 
$S_3$-invariance of the Hamiltonian.  For the desired count 
of the gapless modes it suffice to maintain intact the 
structure of $S_2$ angular harmonics. 
The lifting of a model to an entire  $S_3$ invariance, which can be done along the lines of \cite{Gia2}, is unnecessary for the present purposes.   \\

 The, above Hamiltonian possess an infinite family of 
macro-states labeled by the occupation number of the 
corresponding master 
modes $n_{0_N} = N$ and the energies $E_N \simeq N\epsilon_{0_N}$, where $N=(s+1)^2$.  
On each of these states 
the corresponding modes $a_{slm}$ become gapless. 
For large $s$, their number is $N \sim s^2$. The entropy of the degenerate  
micro-states created by these gapless modes therefore scales as $\sim s^2$.  This represents an area of a sphere of radius $\sim s \sim \sqrt{N}$ measured in units of some fixed fundamental length. \\

However, the area scaling of entropy {\it a priory} does not specify its energy dependence.  This is an additional 
information encoded in $N$-dependence of the gaps
$\epsilon_{0_N}$.  
 For example, as explained in the previous chapter, the choice $\epsilon_{0_N} = {\Lambda \over  \sqrt{N}}$ reproduces the dependence between the black hole 
entropy and its energy, with $\Lambda$ playing a role of the Planck mass. \\ 

 Thus,  once the neurons are visualized as the angular momentum modes of a field, the above network
exhibits a ``holographic" behaviour of the type 
\cite{Gia3}. Here the term {\it holography} is manifested in the fact that  the emergent gapless modes effectively ``inhabit" an area of a lower dimensional sphere.
 This is intriguing, since holography  
is usually considered to be 
characteristic of the gravitational systems \cite{Hol1, Hol2,Hol3,Hol4,Hol5}.  \\
  
 Applying the previous analysis to the above Hamiltonian, we reach the following conclusion.  Starting from an arbitrary initial state 
 $\ket{in}=\ket{1}_{b_N}$ with a single excited quantum of one of 
 the $b_N$-modes, for arbitrarily large $N$, the  Hamiltonian generates 
 an unsuppressed transition into a classical state with 
 $n_{0_N} =N=(s+1)^2$, with the micro-state entropy scaling 
 as  $\sim s^2$.   \\
 
 It is straightforward to further enhance the complexity of the network, and allow for more channels of classicalizing transitions.  
 We shall not enter this model building here, since the presented example suffice for explaining the key ideas
 of the framework.

\subsection{Hierarchy Problem: Standard Model as Brain Network} 

 Our analysis can provide a guideline for UV-completion
 of the Standard Model through the mechanism of classicalization of the Higgs field. 
   For accomplishing this goal, the Higgs boson must be endowed by a new interaction that would enable its unsuppressed  
 transition into some multi-particle states \cite{Cl1}.  The lesson we have learned is that such states must posses 
 an exponentially enhanced
 memory storage capacity. This is necessary for 
 compensating the exponential suppression of the transition amplitude. \\

 A quantum  neural network of the 
 type (\ref{HHH}, \ref{AH}) represents a toy effective model describing the 
 classicalization of the Higgs field in the Standard Model. 
 In order to see this, we should perform the following identification. 
  First of all, we must identify the neural degrees of freedom with the momentum modes of the quantum fields.  \\ 
 
 In order to understand who is who, let us recall \cite{Cl1} that the key ingredient of solving the hierarchy problem via classicalization is to allow an unsuppressed transition of the high energy Higgs particles into a classical state
 composed out of many soft quanta. \\

  Thus, we establish the following dictionary between 
  the classicalizing neural network and the Standard Model.  
 The hard $b$-neurons of energy $N\epsilon_0$ 
 must be mentally replaced by the high momentum modes of the Higgs field with the center of mass energy given by 
 $N\epsilon_0$. For example,  $\hat{b}_N^{\dagger}  \rightarrow  \hat{h}_{\vec{k}}^{\dagger} 
 \hat{\bar{h}}^{\dagger}_{-\vec{k}}$, where $\vec{k}$ is a momentum vector, 
 and $\hat{h}_{\vec{k}}^{\dagger}$ and 
 $\hat{\bar{h}}^{\dagger}_{-\vec{k}}$ denote the creation operators of the Higgs boson and of its antiparticle respectively.  \\
 
\begin{figure}
 	\begin{center}
        \includegraphics[width=0.53\textwidth]{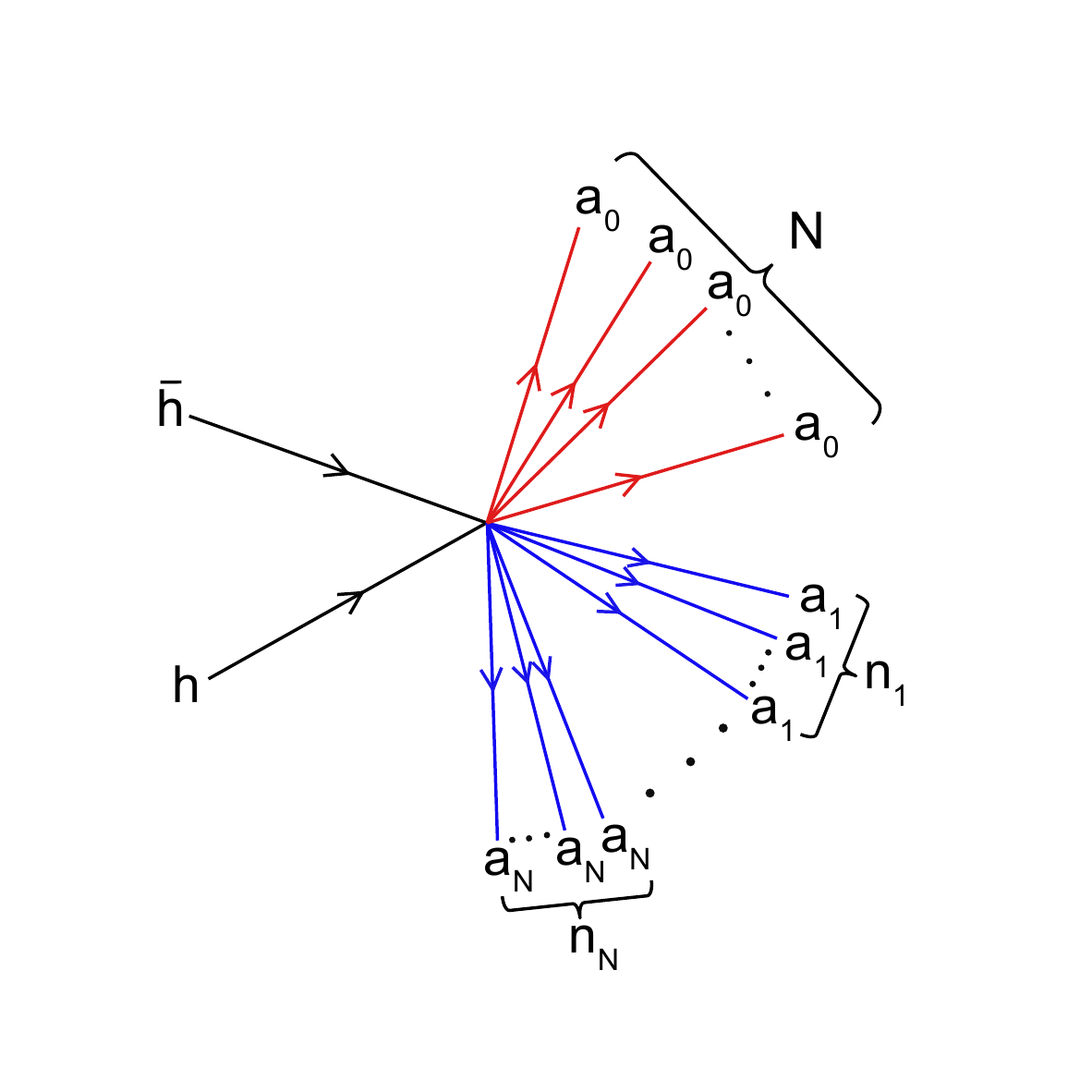}
 		\caption{The Feynman diagram representation 
	of the classicalizing transition described in the text. 	
 The ingoing and outgoing arrows describe the destroyed 
 and created particles respectively.
 The black lines denote an incoming pair of Higgs
 and anti-Higgs particles. The red lines 
 denote the outgoing {\it soft} $a_0$-quanta and  
the blue lines denote the outgoing {\it super-soft}  $a_{k \neq 0}$-particles respectively.  For each particular set of $n_a,...n_N$ the transition rate is exponentially suppressed, but the suppression disappears after summation over the sets.} 	
\label{Feynman}
 	\end{center}
 \end{figure}

    With such a replacement the effective Hamiltonian 
 (\ref{AH}) shall generate classicalizing transitions 
 of high energy Higgs pairs into multi-particle states. 
Namely,  an initial quantum state with Higgs  particle-anti-particle pair of energy $E = 2|\vec{k}| =
N\epsilon_0$, reprented by a ket-vector,  
 $\ket{in} =\ket{1}_{h_{\vec{k}}}\otimes  \ket{1}_{\bar{h}_{-\vec{k}}}\otimes \ket{0}_a$, 
 gets converted into a classical state, 
  $\ket{final} = \ket{0}_{h_{\vec{k}}}\otimes  \ket{0}_{\bar{h}_{-\vec{k}}}\otimes \ket{N}_{a_0} \otimes \ket{n_1}_{a_{1}} \otimes,...,\otimes \ket{n_N}_{a_{N}}$, 
 in which $N$ soft $a_0$-quanta are present.  
 The Feynman diagram corresponding to this process
 is given in Fig.\ref{Feynman}.
As shown above, the rate of the transition is not exponentially suppressed due to the multiplicity of micro-states.     
 \\

  Notice, the $a$-modes can either correspond to the 
  momentum modes of the Higgs field itself or to 
  some additional Bosonic field introduced 
  in the theory.  
  In other words, the theory could classicalize either due to a new energy-dependent self-interaction 
  of the Higgs field or due to its  interactions with the additional fields from beyond the Standard Model. This is a matter of the model building, which is beyond our present goal. \\  
   
    In conclusion, when viewed in the language of 
 a quantum neural network, the mechanism of  UV-completion of the  Standard Model by classicalization would mean that the Standard Model operates as a 
 {\it brain network}  that exploits an arbitrarily high energy - pumped in it through the hard quanta -
 for bringing itself into a classical state 
of the maximal available memory storage capacity. \\
 
   \section{Outlook} 
   
 In this paper we described an effective theory 
 of classicalization \cite{Cl1, Cl2, 2N, 2ND, ClPH}. 
 While our model represents a 
 fully self-consistent quantum system, in parallel, we  
gave its interpretation \cite{Gia1,Gia2} in terms of  a quantum neural network.  In this description, the momentum modes of particles 
 are identified with the neural degrees of freedom, whereas  their interactions are mapped on synaptic connections between the respective neurons. \\
 
 The key aspect was to highlight 
 the role of the enhancement of micro-state entropy  
 due to the emergence of {\it gapless} modes,
 via the mechanism of  \cite{Gia1,Gia2,Gia3}.
 There it was shown that, in case of gravity-like connections, a macroscopic excitation of one of the 
 soft modes renders a set of modes, connected to the  
 excited ``master" mode, effectively gapless. 
 In a neural network language, this means that 
 the system attains a state in which it acquires  
 a maximal capacity of the memory storage and an ability to respond to super-soft input stimuli. In such states, an exponentially large number of patterns   
 can be stored within an arbitrarily narrow energy gap. \\ 

 In order to avoid the potential problems with terminology,  we gave a careful definition of our criteria
 and explained why the network that gives rise 
 to the gapless neurons via our mechanism  is special as compared to a generic
 quantum Hopfield network.  
  \\

  Next, we studied the question of a classicalizing 
 transition from an initial quantum state into a ``classical" state of $N$ soft quanta, 
 \begin{equation} \label{1N}
   \ket{1}_{\rm quant} \rightarrow  \ket{N}_{\rm class}\,.
   \end{equation}  
 Using a simple effective Hamiltonian, we 
  first explained why the transitions to the individual $N$-particle states must be exponentially-suppressed by the factors  ${\rm e}^{-N}$. \\

  To the best of our knowledge, the presented way of reasoning showing that the 
 physical amplitudes of the processes that increase occupation numbers must be  exponentially suppressed,  
 is new and can be useful on its own right. 
 In particular, this gives yet another non-perturbative  argument showing that without introduction of new classicalizing interactions that become strong above few TeV energies,  in the Standard Model alone the production of classical states with  high occupation number of the Higgs quanta should remain exponentially suppressed
(regardless of a possible breakdown of perturbation 
theory (see, e.g., \cite{breakdown})) and the Hierarchy Problem
 will remain unsolved.   
 \\
        
  We then showed, how this suppression is compensated 
 by the multiplicity of the micro-states in cases when the 
 classical $N$-particle state exhibits an exponentially enhanced memory capacity.   We showed that a
 necessary condition (\ref{NSP}) is that 
 a high occupation number $N$ of each soft 
 master mode
 $a_{0_N}$ composing the classical state,  is matched by the equal number of species ($a_{k_N},\, k=1,2,...N$) of the emergent {\it gapless}  modes. \\

 In such a situation, the production of $N$ soft quanta 
 of $a_{0_N}$-mode 
 is accompanied by a number distribution,  $n_1,n_2,...,n_N$,
 of the super-soft ones $a_{k_N}$, coming from the gapless modes. 
  It is the summation over the distributions of these super-soft quanta that provides the desired 
 exponential enhancement of the transition probability.   
 \\
 
 There is something striking about understanding  this phenomenon in terms of quantum neural networks.    
  In this language, the effect of classicalization is translated as the 
   ability of a network to readily bring itself into a 
 (classical) state of maximal memory storage capacity, 
  in response to an external quantum stimulus of very high energy. \\
  
    The presented way of looking at the 
  field-theoretic systems,
 such as, the classicalizing Standard Model
 Higgs  and/or quantum creation of black holes, 
 puts the seemingly-disconnected phenomena 
in an unified perspective.  They all emerge as the brain networks that aspire to attain the states of maximal available  memory capacity.

\section*{1.\, Note added}

Allen Caldwell shared with us the results of his numerical analysis \cite{Allen}, which show that under a random walk evolution the enhanced memory 
states of the  neural network model of \cite{Gia1} 
are the attractor points to which the evolution converges. 
The Hamiltonian of \cite{Gia1}, used in this analysis, 
represents a slightly expanded version of (\ref{Hamilton}), with unessential difference.  
 It appears that this convergence, primarily due to its stochastic  nature, although seen 
by a different protocol, captures some aspects of the quantum evolution presented here.  This study  would be worth pursuing further.

  \section*{2.\, Note Added} 
  
  After this paper was finished, we became aware of the 
  work by Cesar Gomez \cite{CesarIR} in which he is  also 
  discussing the role of infrared physics in classicalization, but from a different angle.  
  Although Gomez in his analysis is touching neither a mechanism of micro-state entropy nor the emergence of the gapless modes, nevertheless, the clear-cut connection point between the two proposals is the crucial role of dressing by the infrared 
physics in classicalizing theories.
In our approach, this is expressed as an  accompaniment to a large occupation number of each soft mode, by 
an equal number of the emergent  gapless {\it species}. It is the summation over the distributions of the
super-soft quanta of these gapless modes that makes 
an unsuppressed classicalization possible.

\section*{Acknowledgements}
 We thank Allen Caldwell, Cesar Gomez, 
 Tamara Mikeladze-Dvali and Sebastian Zell for discussions. Some thoughtful questions about 
ideas of \cite{Gia1,Gia2} from George Musser are 
also acknowledged. 
This work was supported in part by the Humboldt Foundation under Humboldt Professorship Award, ERC Advanced Grant 339169 "Selfcompletion", by TR 33 "The Dark Universe", and by the DFG cluster of excellence "Origin and Structure of the Universe". 

\appendix


\begin{thebibliography}{10}


	
\bibitem{Cl1} G. Dvali, G.F. Giudice, C. Gomez and A. Kehagias, UV-Completion by Classicalization,
JHEP 1108, 108 (2011) [arXiv:1010.1415 [hep-ph]]; 

For a summary and more complete list of references, see: 
G.~Dvali,  	
Strong Coupling and Classicalization, 
 Subnucl, Ser. 53 (2017) 189-200;  arXiv:1607.07422 [hep-th]. 

\bibitem{Cl2}
G. Dvali, C. Gomez and A. Kehagias, Classicalization of Gravitons and Goldstones,
JHEP 1111, 070 (2011) [arXiv:1103.5963 [hep-th]].

\bibitem{2N} 
G. Dvali, C. Gomez, R.S. Isermann, D. L\"ust, S. Stieberger, 
Black hole formation and classicalization in ultra-Planckian $2\rightarrow N$ scattering, Nucl.Phys. B893 (2015) 187-235, arXiv:1409.7405 [hep-th]; \\

\bibitem{2ND} A.~Addazi, M.~ Bianchi, G.~Veneziano, Glimpses of black hole formation/evaporation in highly inelastic, ultra-planckian string collisions 
JHEP 1702 (2017) 111,  arXiv:1611.03643 [hep-th]

\bibitem{SELF}
G.~Dvali, C.~Gomez, 
Self-Completeness of Einstein Gravity,
arXiv:1005.3497 [hep-th] 



\bibitem{BH1}
G.~t Hooft, Graviton Dominance in Ultrahigh-Energy Scattering, Phys. Lett. B198,
61-63 (1987);

\bibitem{BH2} D. Amati, M. Ciafaloni and G. Veneziano, Superstring Collisions at Planckian Energies,
Phys. Lett. B 197, 81 (1987); Classical and Quantum Gravity Effects from Planckian
Energy Superstring Collisions, Int. J. Mod. Phys. A 3, 1615 (1988); Can Space-Time Be
Probed Below the String Size?, Phys. Lett. B 216, 41 (1989); Higher Order Gravitational
Deflection and Soft Bremsstrahlung in Planckian Energy Superstring Collisions, Nucl.
Phys. B 347, 550 (1990); Effective action and all order gravitational eikonal at Planckian
energies, Nucl. Phys. B 403, 707 (1993).

\bibitem{BH3} D.J. Gross and P.F. Mende, The High-Energy Behavior of String Scattering Amplitudes,
Phys. Lett. B 197, 129 (1987); String Theory Beyond the Planck Scale, Nucl. Phys. B
303, 407 (1988).

\bibitem{Gia1} G.~Dvali,  Critically excited states with enhanced memory and pattern recognition capacities in quantum brain networks: Lesson from black holes,  arXiv:1711.09079 [quant-ph].

	
\bibitem{Gia2} G.~Dvali, Black Holes as Brains: Neural Networks with Area Law Entropy,  arXiv:1801.03918 [hep-th]. 

\bibitem{Gia3} G.~Dvali,  Area Law Micro-State Entropy from Criticality and Spherical Symmetry,   
arXiv:1712.02233 [hep-th], Phys. Rev. D, to appear. \\ 

 \bibitem{Crit} 
   This type of criticality, originally was proposed  
to be the source of black hole entropy,
within the picture of \cite{NP}, in,   
	
G.~Dvali and C.~Gomez, {\em {Black Holes as Critical Point of Quantum Phase
			Transition}\/},  \href{http://dx.doi.org/10.1140/epjc/s10052-014-2752-3}{Eur.
		Phys. J. C {\bf 74} (2014)  2752},
	\href{http://arxiv.org/abs/1207.4059}{{\tt arXiv:1207.4059 [hep-th]}}.

\bibitem{NP}
  G.~Dvali and C.~Gomez,
  Black Hole's Quantum N-Portrait,
  Fortsch.\ Phys.\  {\bf 61} (2013) 742
  [arXiv:1112.3359 [hep-th]].  
	

\bibitem{kak} 
 S.C.~Kak, Quantum Neural Computing, Advances in Imaging and
Electron Physics, 94 (1995) 259-314.


\bibitem{Hopfield} 
J.J. Hopfield, {\em Neural Networks and Physical Systems with Emergent Collective Computational Abilities}, 	Proc, Natl. Acad. Sci. USA 79 (1982) 2554

\bibitem{QNETS} M.~Schuld, I.~Sinayskiy, and 
F.~Petruccione, The quest
for a quantum neural network, Quantum Information
Processing, DOI 10.1007/s11128-014-0809-8, 2014.

\bibitem{Bek}
	J.~D. Bekenstein, {\em Black holes and entropy\/},
	\href{http://dx.doi.org/10.1103/PhysRevD.7.2333}{Phys. Rev. D {\bf 7} (1973)
		no.~8, 2333--2346}.

 \bibitem{largeN} G.~'t Hooft,  ``A Planar Diagram Theory for Strong Interactions",
Nucl. Phys. B72, 461, (1974).

	\bibitem{bogoliubov}
	N.~Bogoliubov, On the theory of superfluidity,  J. Phys {\bf 11}  no.~1(1947) 23 .
			
		

 \bibitem{Hol1} 
 
 G.'t Hooft, Dimensional reduction in quantum gravity, gr-qc/9310026; 
 
 L. Susskind,
The World As A Hologram, J. Math. Phys. 36, 6377 (1995), hep-th/9409089


 \bibitem{Hol2} 
 
 J.M.~Maldacena, The large N limit of superconformal field theories and supergravity,
Adv. Theor. Math. Phys. 2, 231 (1998), hep-th/9711200; 

\bibitem{Hol3} 

S.S.~ Gubser, I.R.~Klebanov
and A.M.~Polyakov, Gauge theory correlators from non-critical string theory, Phys.
Lett. B 428, 105 (1998), hep-th/9802109; 

\bibitem{Hol4}
E.~ Witten, Anti-de Sitter space and holography,
Adv. Theor. Math. Phys. 2, 253 (1998), hep-th/9802150;

\bibitem{Hol5} 
L. Susskind and E. Witten, The holographic bound in anti-de Sitter space,
hep-th/9805114
 
 \bibitem{ClPH} 
  C. Grojean, R.S. Gupta, Theory and LHC Phenomenology of Classicalon
Decays, JHEP 1205 (2012) 114, arXiv:1110.5317 [hep-ph].


\bibitem{breakdown}
	
H. Goldberg, Breakdown of perturbation theory at tree level in theories with scalars, Phys. Lett. B 246, 445 (1990); 

J.M. Cornwall, On the High-energy Behavior of Weakly Coupled Gauge Theories, Phys.
Lett. B 243, 271 (1990).

  	
 

\bibitem{Allen} A.~Caldwell, private communications. 

\bibitem{CesarIR} C.~Gomez,  to be published. 


%
\end{thebibliography}
\end{document}